\documentclass{aa}
\usepackage{graphicx}
\usepackage{txfonts}
\begin{document}
\authorrunning{S.~Kuznetsov et al.}
\titlerunning{Observations of EXO~2030+375 with IBIS/ISGRI}
  \title{First observations of the X-ray transient EXO~2030+375 with
IBIS/ISGRI\thanks{Based on observations with INTEGRAL, an ESA project with instruments and science data centre funded by ESA member states (especially the PI countries: Denmark, France, Germany, Italy, Switzerland, Spain), Czech Republic and Poland, and with the participation of Russia and the USA.}
}

   \author{Sergey Kuznetsov\inst{1}, Maurizio Falanga\inst{1}, Pere
   Blay\inst{2}, Andrea Goldwurm\inst{1}, Paolo Goldoni\inst{1},
         \and Victor~Reglero\inst{2}
          }
   \offprints{S. Kuznetzov\\
\email{kuznetz@cea.fr}
}

   \institute{\inst{1}CEA Saclay, DSM/DAPNIA/Service d'Astrophysique, 91191,  Gif sur Yvette Cedex, France \\
    \inst{2}GACE, Instituto de Ciencia de los Materiales, Universitat de Valencia, P.O. Box 22085, 46071 Valencia, Spain
             }

   \date{July 2003}

   \abstract{We present a first INTEGRAL observation of the 42s
   transient X-ray pulsar EXO~2030+375 with IBIS/ISGRI. The source was
   detected during Cyg X-1 observations in December 2002.  We analyzed
   observations during the outburst period from 9 to 21 December 2002
   with a total exposure time of $\sim770$ kiloseconds. EXO~2030+375
   was almost always detected during single $\sim30$ minute exposures
   in the 18$-$45 energy bands. The source light curve shows the
   characteristic outburst shape observed in this source.
   \keywords{gamma rays: observations -- X-rays: individual:
   EXO~2030+375 -- Telescopes: INTEGRAL: IBIS} }

   \maketitle
%

\section{Introduction}

EXO~2030+375 is a member of the Be/X-ray transients systems, which are
the most common type of accreting X-ray pulsar. They consist of an
accreting neutron star and a B spectral main-sequence donor star that
shows Balmer emission lines (\cite{apparao} for review). The line
emission is believed to be associated with an equatorial outflow of
material expelled from the rapidly rotating Be star that forms a
quasi-Keplerian disk near the Be star ({\cite{quirrenbach}). The X-ray
emission of the transient pulsar EXO~2030+375 is modulated by $\approx
42$ s pulsations and periodic $\approx 46$ days Type I outbursts, that
are produced at each periastron passage of the neutron star, i.e. when
the pulsar interacts with the disk of the Be star.

\noindent EXO~2030+375 was discovered in 1985 May with EXOSAT satellite
during a large outburst phase (\cite{parmar89b}). This outburst was
first detected at a 1$-$20 keV energy band and its luminosity is close
to the Eddington limit (assuming 5 kpc distance to the source) for a
neutron star (\cite{parmar85}). During the later EXOSAT
observations a monotonic decline in intensity was seen over nearly 3
orders of magnitude. During this luminosity decline, the intrinsic
spin period changed dramatically, with a characteristic spin-up
timescale $-P/\dot{P} \approx 30$ years (\cite{parmar89b}), the energy
spectrum (\cite{reynolds}), and the $1-10$ keV pulse profile
(\cite{parmar89a}) all showed significant luminosity dependence. Such
a spin-up indicates the presence of an accretion disk penetrating well
inside the corotation radius.  Further evidence of an accretion disk
resulted from the detection of 0.2 Hz quasi-periodic oscillations
(\cite{angelini}).\\
\noindent The shape of the continuum X-ray spectrum in the range 2$-$25
keV of EXO~2030+375 can be represented by a powerlaw ($\alpha\approx
1.5$) modified at energies above a high energy cutoff
$E_{\mathrm{cut}}\approx 20$keV
with $E_{\mathrm{fold}}\approx 30$keV (\cite{reynolds}). 
Evidence of a possible cyclotron feature at 36 keV was found in
spectra with RXTE observations (\cite{reig}).\\
\noindent We report first observation results of EXO~2030+375 made
with INTEGRAL/ISGRI during the Performance and Verification (PV)
phase.  The source was observed for more than 10 days in different
offaxis position.

\section{Instruments and Observations}

\noindent The International Gamma-Ray Astrophysics Laboratory
(INTEGRAL) (\cite{winkler}) is a 15 keV$-$10 MeV
gamma-ray observatory with concurrent source monitoring at X-rays
(3$-$35 keV) and in the optical range (V, 500$-$600 nm). 
The INTEGRAL main gamma-ray instruments are the
spectrometer SPI and the imager IBIS, the supplementary instruments are
the X-ray monitor JEM-X and the Optical Monitoring Camera OMC.
This provide a combination of imaging and spectropy over a wide range
of X-ray and gamma-ray energies including optical monitoring.

\noindent The imager IBIS with a angular resolution of 12\arcmin\ FWHM,
a wide full coded field of view (FOV) of $9\degr\times9\degr$ and a
partially coded FOV of $29\degr\times29\degr$ (\cite{ubertini})
 consists of two detection layers, ISGRI and PICsIT. The front
layer ISGRI (\cite{lebrun}) is sensitive for lower energy
(15 keV$-$1 MeV), and while its peak sensitivity is between 15 keV
and 200 keV, the second one PICsIT (\cite{di}) is sensitive between
$\sim$200 keV and $\sim$8 MeV. The division into two layers allowed
the paths of the photons to be tracked in 3D, as they scatter and
interact with more than one element. This offers the possibility to
operate IBIS in the additional ``Compton mode''.  The present results
of the analyzed data are produced by the ISGRI detector layer.

\noindent The Cygnus region, including in the field of view Cyg~X-1
(Laurent et al. this volume), Cyg~X-3 (\cite{goldoni})
and EXO~2030+375 was observed 15 November$-$21 December 2002. The
main target, observed in different position, was Cyg~X-1
(\cite{bazzano}). During this period EXO~2030+375 was in the
IBIS/ISGRI field of view and detected in the outbursts, i.e., in the
periastron passage phase starting from December 9th, 2002. In this
Letter we report on results of analysis of EXO~2030+375 observations
carried out by IBIS/ISGRI in 9$-$21 December 2002.

\section{Data Reduction}

\noindent During the PV phase of observations, the instrument
parameters and moreover the pointing direction frequently changed and
EXO~2030+375 was not fully coded during all analysed observations.  We
used observational data starting from December 9th, 2002 when source
appeared in the X-ray image of Cygnus region and was reliably
detected by IBIS/ISGRI. We rejected observations carried out in slew mode
and those in which PICsIT was operated in ``Photon-by-Photon'' mode.  

\noindent For light curve data we analyzed observations starting with 
the outburst phase of EXO~2030+375 from 9 December and lasting to 21
December 2002. These observations correspond to INTEGRAL orbit
revolutions 19$-$22. Table~\ref{tab1} summarizes the observations time and
mean source flux for each revolution.

  \begin{table}[htb]
      \caption[]{Observation log and average fluxes of EXO~2030+375 in the period of 9$-$21 December 2002. }
         \label{tab1}
     $$ 
         \begin{array}{ccccc}
            \hline
            \hline
            \noalign{\smallskip}
             \rm Revolution & \rm Start-Stop & \rm Total\ exp.,  & \rm Eff.\ exp., & 18-45\rm\ keV\\
            \noalign{\smallskip}
            \rm number & \rm time, MJD & \rm ksec  &  \rm ksec & \rm flux, cnt/s\\
            \noalign{\smallskip}
            \hline
            \noalign{\smallskip}
      19 & 52617.3-52620.1 &  192 & 142 & 5.42\pm0.05 \\
      20 & 52620.4-52623.0 &  179 & 56  & 7.82\pm0.08 \\
      21 & 52623.3-52626.1 &  185 & 111 & 6.50\pm0.07 \\
      22 & 52626.3-52629.0 &  214 & 165 & 2.63\pm0.07 \\
            \noalign{\smallskip}
            \hline
         \end{array}
    $$ 
\end{table}

\noindent The images produced by the telescope were analysed using the ISDC
public software and software developed at our institute for
calibration (INTEGRAL Off-line Scientific Analysis (OSA) v1.1 with
latest versions of ii\_skyimage~3.5 and ii\_spectra\_extract~1.9).
The images were deconvolved using the procedures described in
\cite{goldwurm} to recover source position and flux.

\section{Imaging and Light Curve}

\noindent Cyg~X-1 is one of the brightest hard X-ray sources in the sky.
It was the main target of the IBIS/ISGRI pointing observations in
December 2002 of the Cygnus region. EXO~2030+375 and Cyg~X-3 are two
other X-ray sources in the vicinity of Cyg~X-1 (see Fig.~1). All three
sources were in the IBIS/ISGRI field of view during monitoring of this
part of X-ray sky. The effective coded area of source on the detector
plane decreased consequently from $\sim80\%-100\%$ in revolution 19 to
$\sim50\%-80\%$ in revolution 22.

  \begin{figure}[htb] \centering \includegraphics[width=9 cm]{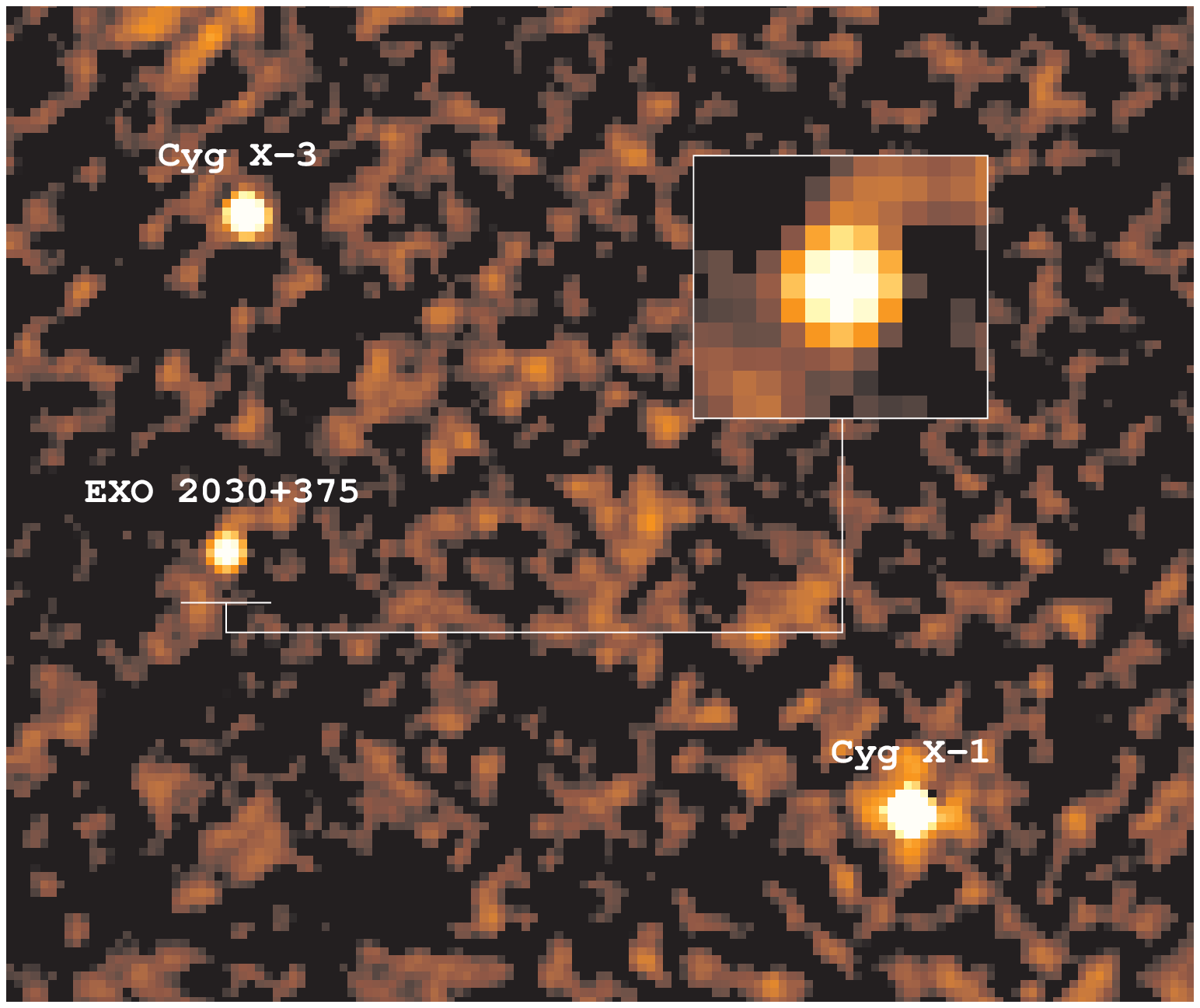}
  \caption{Image of Cygnus region in 18-45 keV energy band averaged
  over 142 ksec of observations during 19th revolution. Image size is
  $\sim9\fdg7\times11\fdg7$, size of zoomed region of
  EXO~2030+375 centered at $20^{\mathrm h} 32^{\mathrm m} 19\fs8
  +37\degr 38^{\mathrm m} 15\fs5$ is
  $0\fdg9\times0\fdg9$. Sources were detected with following
  significance of $1300\sigma$, $210\sigma$ and $90\sigma$ for
  Cyg~X-1, Cyg~X-3 and EXO~2030+375, respectively. Squared root scale
  was used for color representation.} \label{figimg} \end{figure}


   \begin{figure*} \centering \includegraphics[width=11 cm, angle=
   -90]{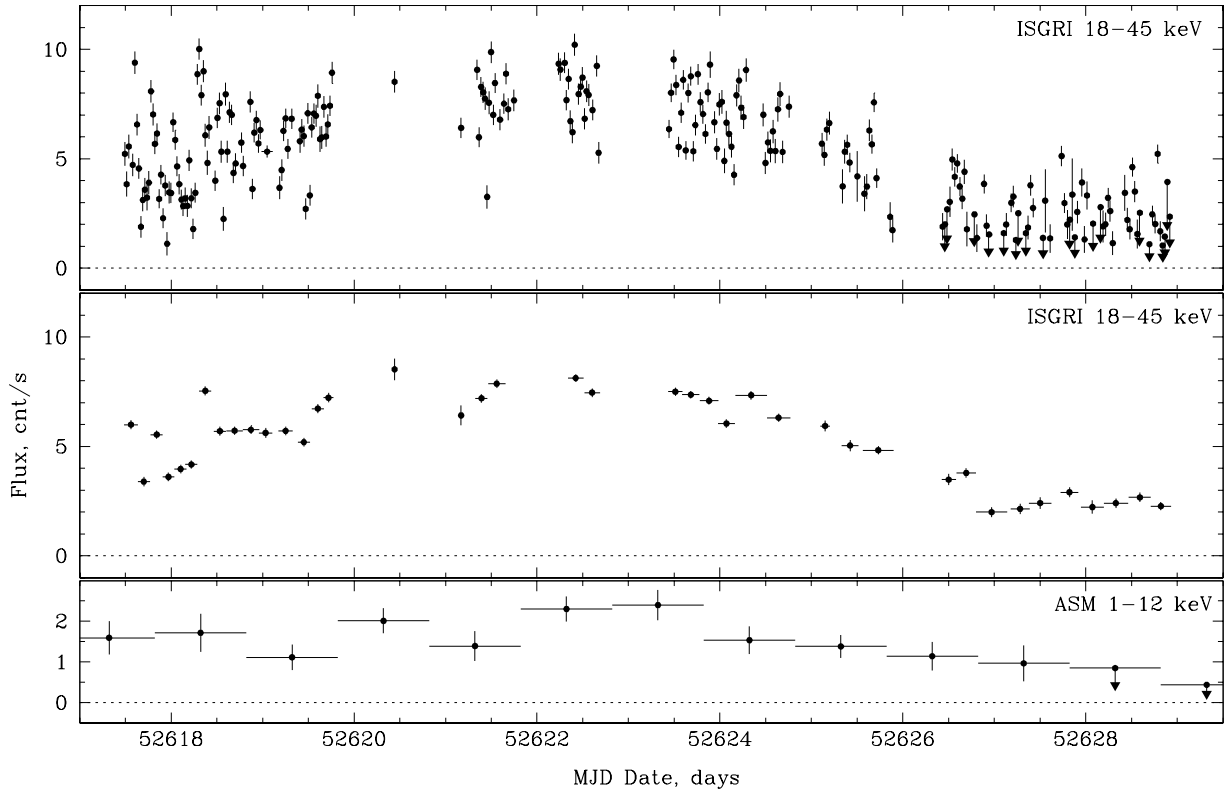} \caption{Light curves of EXO~2030+375 in 18$-$45 keV
   (top and middle) and 1$-$12 keV energy bands. Data points of top
   pannel correspond to observations of each particular ``Science
   Window'' and count rate was extracted directly from source
   reconstructed spectra.  Source flux below 2$\sigma$ detection level
   replaced by 2$\sigma$ upper limits.  Data averaged over 1$-$6
   ``Science Windows'' are shown in middle panel.  Light curve
   obtained bt RXTE All Sky Monitor (ASM) is shown in bottom for
   comparison. Correlation of source fluxes between in IBIS/ISGRI and
   RXTE/ASM observations is apparent.} \label{figlc}%
\end{figure*}

\noindent Before the image reconstruction  and the extraction of light
curve from the data we studied the source intensity in different
energy bands. Spectral shape of EXO~2030+375 is well approximated by a
power-law spectrum with a high energy cutoff
(\cite{reynolds}). Therefore we preferred to select softer energy
channels than harder ones in order to avoid possible contamination of
the source flux by background radiation. For revolutions 19$-$22 the
level of background radiation is not well established or at least not
as well as for later observations when parameters of ``veto'' were
fixed. We found that both observational parameters of EXO~2030+375 --
count rate and significance of the detection level of source flux have
maximum in 18$-$45 keV energy range. We used this energy band over our
analysis.

\noindent The image of the Cygnus region in 18$-$45 keV energy range averaged
over all ``Science Window'' observations carried out in the course of
the revolution 19 is shown in figure~1.  We detected Cyg~X-1 with
unprecedented significance of $\sim1300\sigma$ level. Its X-ray flux
dominated over Cyg~X-3 and EXO~2030+375 fluxes which were detected
with significance levels $\sim10-20$ times lower (see Fig.~1) than
Cyg~X-1.

\noindent In order to get EXO~2030+375 light curve in 18$-$45 keV we extracted
source count rate from deconvolved image of every ``Science Window''
observation. The image analysis pipeline allows to get and estimate
intensity of interesting sources directly in output of files.
Unfortunately EXO~2030+375 flux extracted in a such way was found at
too low intensity level in some part of observational data. In order
to check the flux determination procedure we extracted source count
rate directly from source reconstructed spectra and compared results
of both methods. It was found that extraction of count rate from
source spectrum is a more reliable and stable method than the
extraction from images. The problem was due to the fact that the
detector's count rate is dominated by very strong hard X-ray source
like Cyg~X-1.

\noindent Light curve of EXO~2030+375 is shown in figure~2. In order
to compare IBIS/ISGRI results with results obtained with other
observatories we plotted RXTE/ASM data in the bottom of figure~2. The
correlation between averaged 18$-$45 keV IBIS/ISGRI flux and 1$-$12 keV
flux of RXTE/ASM is apparent. The maximum intensity of outburst of
$\sim8-10$ cnt/s ocurred during MJD $\sim52620-52624$. This outburst
was also detected by JEM-X (\cite{martinez}).

The time range of the first and the last observed photon event correspond to 
the  orbital phase of EXO 2030+375 from 0.9849 to 0.2332.  The periastron 
passage was at the orbital phase zero at 2452618.683 JD  $\sim$ 2.8 days before
the outburst maximum, according to the epemeries given by \cite{wilson}.

\section{Spectral Analysis}

\noindent Only fully coded observations of EXO~2030+375 were used to
analyze spectral shape. Due to some uncertainties in calculations of
counts in spectral bins in partially coded observations, such data
were not used for averaged spectrum (nevertheless we used them for
light curve shown in figure~2). Due to absence of fully coded
observations of EXO~2030+375 in revolutions 20$-$22 spectral analysis
was not performed using these revolutions.

Spectral extraction was performed for each particular ``Science
Window''. Source spectra of 31 ``Science Window'' of fully coded
observations of EXO~2030+375 with total integration time of $\sim 49$\
ksec were averaged in 13 channels rebinned from 2048 channels of the
original matrix. We fitted the intensities of the background and the 3
IBIS/ISGRI sources (Cyg~X-1, Cyg~X-3 and EXO~2030+375) in the field of
view for each spectral bin. The averaged spectrum of EXO~2030+375 is
shown in figure~3.

We used the 22$-$100 kev energy range for spectral fit. Power-law model with
exponential cutoff was used to approximate the spectral data. It can
be seen from the figure~3 that the 2nd spectral bin corresponding to the
$\sim25-30$ keV energy band deviates from the expected spectral
shape. This bin was excluded from spectral analysis as the observed
feature is due to a known problem in the present correction tables
used to convert event amplitude channels in deposited energies.  In
order to demonstrate it we present the ratio of the EXO~2030+357 spectrum to
the Crab one in the bottom of figure~3. As it can be seen from figure~3, the
spectral shape of EXO~2030+375 is relatively smooth in comparison to the
Crab spectrum and there are no sharp change in the data points.

Averaged EXO~2030+375 spectrum was well fitted with the model
mentioned above. The following results were obtained: power-law
spectral index $\alpha=1.5\pm0.2$, $E_{\mathrm{cut}}=37\pm4$,
$E_{\mathrm{fold}}=32\pm4$, $\chi^{2}_{\mathrm{red}}=1.09$. These spectral parameters
are close to those obtained by Reynolds et al. (1993).

  \begin{figure} \centering \includegraphics[width=9 cm]{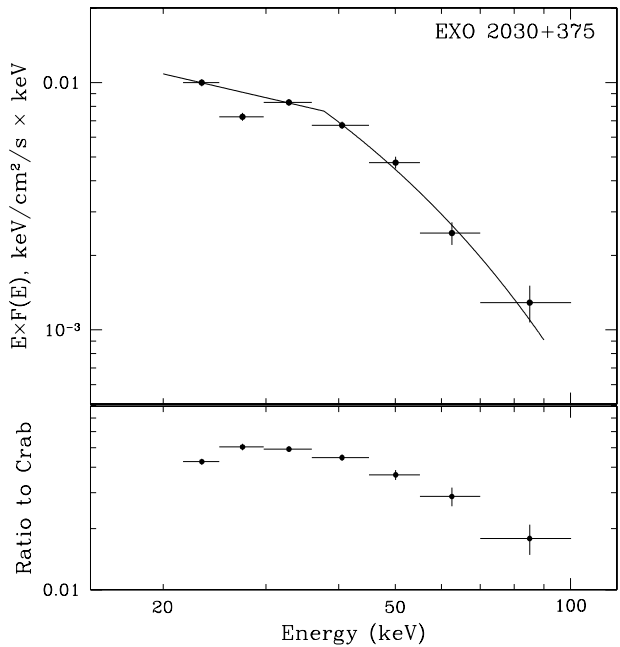}
  \caption{E-folded spectrum of EXO~2030+375 in 20$-$100 keV energy
  range averaged over 31 science windows of revolution 19. Only images
  for which the source was fully coded were used for sum. Total
  accumulated time is $\sim49$ ksec. Spectral fit by power-law model
  with high energy cutoff is shown by the solid line. Ratio to
  e-folded spectrum of Crab is shown for comparison in the bottom.}
  \label{} \end{figure}

\section{Conclusions}

\noindent We present a first INTEGRAL observation of the transient X-ray
pulsar EXO~2030+375 with IBIS/ISGRI. We demonstrate that the
results on EXO~2030+375 are consistent with those obtained from
other X-ray missions. Source light curve shows typical changes
in intensity which correspond to the source state in outburst phase.
The spectrum of EXO~2030+375 in 20$-$100 keV energy band is a power-law with
a spectral index $\alpha=1.5$ and has a high energy cutoff above
$\sim30-40$ keV.

This letter illustrates the IBIS/ISGRI detector capability to produce
scientific results event for partially coded off-axis sources.  This
capability is one of the most important IBIS/ISGRI characteristics and
its exploitation will be very important during the mission.


\end{document}